\documentclass[12pt]{article}
\usepackage{amssymb}
\usepackage[dvips]{epsfig}

\newcommand{\be}{\begin{equation}}
\newcommand{\ee}{\end{equation}}
\def\bea{\begin{eqnarray}}
\def\eea{\end{eqnarray}}

\newcommand{\bn}{\begin{eqnarray}}
\newcommand{\en}{\end{eqnarray}}

\newcommand{\p}{\partial}

\newcommand{\nn}{\nonumber}

\newcommand{\no}{\noindent}

\def\bea{\begin{eqnarray}}
\def\eea{\end{eqnarray}}

\newcommand{\beq}{\begin{eqnarray}}
\newcommand{\eeq}{\end{eqnarray}}
\begin{document}

\title{\textbf{A note on the nonuniqueness of the massive Fierz-Pauli theory and spectator fields}}
\author{D. Dalmazi\footnote{dalmazi@feg.unesp.br} \\
\textit{{UNESP - Campus de Guaratinguet\'a - DFQ} }\\
\textit{{Avenida Doutor Ariberto Pereira da Cunha, 333} }\\
\textit{{CEP 12516-410 - Guaratinguet\'a - SP - Brazil.} }\\}
\date{\today}
\maketitle

\begin{abstract}

It is possible to show that there are three independent families
of models describing a massive spin-2 particle via a rank-2
tensor. One of them contains the massive Fierz-Pauli model, the
only case described by a symmetric tensor. The three families have
different local symmetries in the massless limit and can not be
interconnected by any local field redefinition. We show here
however, that they can be related with the help of a decoupled and
non dynamic (spectator) field. The spectator field may be either
an antisymmetric tensor $B_{\mu\nu}=-B_{\nu\mu}$, a vector
$A_{\mu}$ or a scalar field $\varphi$, corresponding to each of
the three families. The addition of the extra field allows us to
formulate master actions which interpolate between the symmetric
Fierz-Pauli theory and the other models. We argue that massive
gravity models based on the Fierz-Pauli theory are not expected to
be equivalent to possible local self-interacting theories built up
on the top of the two new families of massive spin-2 models.

The approach used here may be useful to investigate dual
(nonsymmetric) formulations of higher spin particles.

\end{abstract}

\newpage

\section{Introduction}

Speculations about a possible massive gravity theory have been
raised long ago. In particular, the problems of mass discontinuity
and the appearance of ghosts have been pointed out in
\cite{vdv,zak} and \cite{bd} respectively. In the last years the
interest in massive gravity has been increased, see e.g. the
review work \cite{h12} and references therein. Those works are
driven both by the accelerated expansion of the universe and more
recently by the discovery \cite{rg,rgt,hr} of appropriate mass
terms in spin-2 theories which furnish a correct counting of
degrees of freedom. The fact that those works are all based on the
usual massive Fierz-Pauli (FP) \cite{fp} theory, described by a
symmetric rank-2 tensor, impels us to search for other
descriptions of massive spin-2 particles.

In particular, a weak field expansion in a frame-like formulation
of gravity $e_{\mu \, a} = \eta_{\mu\, a} + h_{\mu\, a} $
naturally leads to a nonsymmetric field $h_{\mu \, a} \ne h_{a\,
\mu}$. The general case of a second order (in derivatives)
Lagrangian for an arbitrary rank-2 tensor $e_{\mu\nu}$ has been
investigated  in the past in \cite{rivers,barnes,bw,pvn73,cko}.
They conclude that the massive FP theory is the only possibility
which avoids ghosts. It turns out that in \cite{ms} and \cite{nfp}
other possibilities have been found which have motivated our
previous work \cite{rank2} where we revisit the classification of
all possible (second-order) descriptions of a massive spin-2
particle in $D=3+1$. We conclude that there are three ghost free
one-parameter family of solutions. In two of those families the
auxiliary\footnote{We use basically the same notation of
\cite{rank2}, in particular, $\eta_{\mu\nu}=diag(-,+,+,+)$, $\,
e_{(\alpha\beta)}=\left( e_{\alpha\beta} +
e_{\beta\alpha}\right)/2$ and $e_{\left[\alpha\beta\right]}=\left(
e_{\alpha\beta} - e_{\beta\alpha}\right)/2$.} fields
$e_{[\mu\nu]}$ are required, only in the FP family there is a
special case with a purely symmetric tensor $ e_{(\mu\nu)} $ with
only one auxiliary field, the trace $e=\eta^{\mu\nu}e_{\mu\nu}$.
In the next sections the results of \cite{rank2} are confirmed in
a rather simple way and the connection of the new models with the
symmetric FP theory is clarified by means of interpolating master
actions \cite{dj}.

\section{Three families}

In \cite{rank2} we have considered a general Lorentz covariant
second order quadratic action for a rank-2 tensor $e_{\mu\nu}$
with 10 arbitrary real constants. Requiring that the propagator
contains only one massive pole in the spin-2 sector, with positive
residue (no ghost), we have obtained, up to the field
redefinitions $e_{\mu\nu} \to e_{\mu\nu} + a\,  \eta_{\mu\nu}\, e
$ and $e_{\mu\nu} \to A\, e_{\mu\nu} + (1-A) e_{\nu\mu}$, three
one-parameter families of models which are displayed in
(\ref{lfpd}), (\ref{lnfp}) and (\ref{la1}). All the three families
lead on shell to the Fierz-Pauli conditions:

\bea e_{\left[\alpha\beta\right]} &=& 0 \quad , \label{c1}\\
e &=& 0 \quad , \label{c2}\\
\p^{\mu}e_{\mu\nu} &=& 0 \quad , \label{c3} \\
(\Box - m^2)e_{\mu\nu} &=& 0  \quad , \label{kg} \eea

The first family depends on the arbitrary real constant $d_-$:

\be {\cal L}_{FP}(d_-) = {\cal L}_{FP}[e_{(\alpha\beta)}] + d_- \frac{m^2}2 e_{\left[\alpha\beta\right]}^2 \quad
. \label{lfpd} \ee

\no and contains the usual ($d_-=0$) massive FP theory:

\be {\cal L}_{FP}[e_{(\alpha\beta)}] = -\frac 12
\p^{\mu}e^{(\alpha\beta)}\p_{\mu}e_{(\alpha\beta)} + \frac 14
\p^{\mu}e\p_{\mu}e + \left\lbrack
\p^{\alpha}e_{(\alpha\beta)}-\frac 12 \p_{\beta} e \right\rbrack^2
-\frac{m^2}2 \left\lbrack e_{(\alpha\beta)}^2 - e^2 \right\rbrack
\, .  \label{lfp} \ee

\no The massless limit of ${\cal L}_{FP}(d_-)$ is invariant under

\be \delta \, e_{\mu\nu} = \p_{\nu} \xi_{\mu} + \Lambda_{\mu\nu} \quad . \label{s1} \ee

\no  where $\Lambda_{\mu\nu} = - \Lambda_{\nu\mu} $. We remark
that although $d_-$ is completely arbitrary in the free theory
(\ref{lfpd}), it has been argued in \cite{zino}, based  upon a
Stueckelberg-like formulation, that one should fix $d_-=1$.

The second family of models \cite{nfp} depends on the free
parameter $c$ and is given by

\bea {\cal L}_{nFP}(c) &=& -\frac 12
\p^{\mu}e^{(\alpha\beta)}\p_{\mu}e_{(\alpha\beta)}  + \frac 16
\p^{\mu}e\p_{\mu}e + \left\lbrack
\p^{\alpha}e_{(\alpha\beta)}\right\rbrack^2 \nn \\
&-&   \frac 13 \left\lbrack \left( \p^{\alpha}e_{\alpha\beta}\right)^2 + \left(
\p^{\alpha}e_{\beta\alpha}\right)^2\right\rbrack - \frac{m^2}{2}(e_{\mu\nu}e^{\nu\mu} + c \,  e^2)
\label{lnfp}\eea

\no The acronym nFP stands for non Fierz-Pauli, since the mass
term does not need to fit in the Fierz-Pauli form ($c=-1$).
Analogous to ${\cal L}_{FP}(d_-)$, the massless limit of ${\cal
L}_{nFP}(c)$ describes a massless spin-2 particle, see \cite{cmu}
and \cite{nfp}, it is invariant under linearized
reparametrizations plus Weyl transformations

\be \delta \, e_{\mu\nu} = \p_{\nu} \xi_{\mu} + \eta_{\mu\nu}\phi
\quad . \label{s2} \ee

\no The Weyl symmetry can be extended to the whole massive theory
if we choose $c=-1/4$, in which case we get rid of the trace
$e=\eta^{\mu\nu}e_{\mu\nu}$ such that we only have $e_{[\mu\nu]}$
as auxiliary fields.

The third and last family depends on the arbitrary real constant
$a_1$ introduced in \cite{rank2},

\bea {\cal L}_{a_1} &=& -\frac 12
\p^{\mu}e^{(\alpha\beta)}\p_{\mu}e_{(\alpha\beta)} + \left( a_1 +
\frac 14 \right) \p^{\mu}e\left\lbrack \p_{\mu}e - 2
\p^{\alpha}e_{(\alpha\mu)}\right\rbrack + \left\lbrack
\p^{\alpha}e_{(\alpha\beta)}\right\rbrack^2 \nn \\
&+& \left( a_1 - \frac 14 \right) \left(
\p^{\alpha}e_{\alpha\beta}\right)^2 -
\frac{m^2}{2}(e_{\mu\nu}e^{\nu\mu} - e^2) \quad . \label{la1} \eea

\no Differently from ${\cal L}_{FP}(d_-)$ and ${\cal L}_{nFP}(c)$,
the massless limit of ${\cal L}_{a_1}$ describes a massless spin-2
particle plus a massless scalar field (scalar-tensor). The
massless theory is unitary if $a_1 \le -1/12$ or $a_1 \ge  1/4$,
see \cite{rank2}. At $a_1=1/4$ the family ${\cal L}_{a_1}$ becomes
the massive FP theory with $d_-=1$ while at $a_1=-1/12$ it becomes
${\cal L}_{nFP}(c=-1)$. Those are the only intersecting points of
${\cal L}_{a_1}$ with the other two families.

The massless limit of ${\cal L}_{a_1}$ is invariant only under
linearized reparametrizations in general,

\be \delta \, e_{\mu\nu} = \p_{\nu} \xi_{\mu} \quad , \label{s3}
\ee

\no except at $a_1=1/4$ where the massless symmetry is enlarged by
antisymmetric shifts as in (\ref{s1}) and at $a_1=-1/12$ where a
Weyl symmetry shows up as in (\ref{s2}). The case $a_1=-1/4$
corresponds to the model of \cite{ms} where it has been coupled to
a maximally symmetric background. In the next section we
interconnect the models (\ref{lnfp}) and (\ref{la1}) with the
symmetric massive FP theory ${\cal L}_{FP}(d_-=0)$ via master
actions.

\section{Master actions}

There is no local field redefinition which relates  (\ref{lnfp})
or (\ref{la1}) to the symmetric massive FP model. The difficulty
lies in the presence of the antisymmetric tensor $e_{[\mu\nu]}$ in
the derivative terms of (\ref{lnfp}) and (\ref{la1}). However,
since all the three families have the same particle content (for
nonzero mass) one should be able to interconnect them somehow.
Since the FP family (\ref{lfpd}) is obtained from the usual FP
theory by the addition of a pure (arbitrary) mass term for a
decoupled non dynamic field $e_{[\mu\nu]}$, we might try to add to
the symmetric FP model $(d_-=0)$ a mass term for some other field.
This is what we do in the next two subsections.

\subsection{Scalar Spectator}

First, let us add a scalar field and define

\be {\cal L}_b = {\cal L}_{FP}[h_{\alpha\beta}] - b\, m^2 \, \varphi^2 + h_{\alpha\beta}T^{\alpha\beta} \quad .
\label{lb} \ee

\no where ${\cal L}_{FP}[h_{\alpha\beta}]$ is given in (\ref{lfp})
and $h_{\alpha\beta} = h_{\beta\alpha}$ is some symmetric tensor.
We have also added a symmetric external source $T^{\alpha\beta}$.
The additional decoupled mass term does not change the particle
content of the massive FP theory. After a shift with an arbitrary
real constant $s$:

\be h_{\mu\nu} \to h_{\mu\nu}  - s \eta_{\mu\nu} \varphi - \frac{2\, s}{m^2} \p_{\mu}\p_{\nu} \varphi \quad .
\label{sh1} \ee

\no The Lagrangian ${\cal L}_b$ becomes

\bea {\cal L}_b &=& {\cal L}_{FP}[h_{\alpha\beta}]  + m^2(6\, s^2-b) \, \varphi^2 -
 3\, s^2 \p^{\mu}\varphi\p_{\mu}\varphi - 3 \, s\, m^2 \varphi \, h
 \nn \\
&+& h_{\alpha\beta}T^{\alpha\beta} - s\, \varphi \, T - (2\, s/m^2) \varphi \, \p_{\mu}\p_{\nu} T^{\mu\nu} \quad
. \label{lb1} \eea

\no The shift (\ref{sh1}) is defined by requiring that derivative
couplings between $\varphi$ and $h$ vanish. Introducing an
auxiliary vector field and integrating by parts  we can rewrite
(\ref{lb1}) in a first order form

\bea {\cal L}_b &=& {\cal L}_{FP}[h_{\alpha\beta}]  +  3 \, s^2
m^2 A^{\mu}A_{\mu} + m^2(6\, s^2-b) \, \varphi^2 +
h_{\mu\nu}T^{\mu\nu}
\nn\\
&-& s \, \varphi\left( 6\, s\, m \, \p \cdot A + 3 \, m^2 h + T + \frac 2{m^2} \p_{\mu}\p_{\nu}T^{\mu\nu}
\right) \label{lb2} \eea

%\bea {\cal L}_{FP}\left\lbrack h_{\mu\nu} + s \eta_{\mu\nu} \varphi + t \p_{\mu}\p_{\nu} \varphi \right\rbrack
%&=& {\cal L}_{FP}[ h_{\mu\nu}] + 3\, s(s-t\, m^2) \p^{\mu}\varphi\p_{\mu}\varphi \nn \\
%&=& 6\, m^2 \, s^2 \, \varphi^2 +(t\, m^2 -2\, s) \p^{\mu}(\p^{\alpha}h_{\alpha\mu}-\p_{\mu}h) + 3 \, s\,
%m^2\varphi \, h  \quad . \label{i1} \eea

\no Due to the specific form of the usual Fierz-Pauli mass term in
${\cal L}_{FP}(d_-=0)$ it is possible to generate a Maxwell
Lagrangian by making another shift in ${\cal L}_b$ and using the
identity

\be {\cal L}_{FP}\left\lbrack h_{\mu\nu} + r\, (\p_{\mu}A_{\nu} + \p_{\nu}A_{\mu})\right\rbrack = {\cal L}_{FP}
[h_{\mu\nu}] - \frac{m\, r^2}2 F_{\mu\nu}^2(A) + 2\, m^2\, r\, A^{\mu}(\p^{\alpha}h_{\alpha\mu}-\p_{\mu}h)
\label{i2} \ee

\no If we choose $r=-s/m$ we cancel out the $\p \cdot A$ term in
(\ref{lb2}). We can bring an antisymmetric field $B_{\mu\nu}$ into
the game by rewriting the Maxwell term in a first order form. We
end up with a master Lagrangian which now involves three extra
fields $(\varphi,A_{\mu},B_{\mu\nu})$ besides $h_{\mu\nu}$:

\bea {\cal L}_{M1} &=& {\cal L}_{FP}[h_{\mu\nu}] + 3\, m^2 s^2
A^{\mu}A_{\mu} - 2\, m\, s \, A^{\mu}\left(
\p^{\alpha}B_{\alpha\mu} + \p^{\alpha}h_{\alpha\mu} - \p_{\mu} h -
\frac{\p^{\alpha}T_{\alpha\mu}}{m^2} \right) \nn \\ &+&
\frac{m^2}2 B_{\mu\nu}^2  + m^2(6\, s^2 - b)\varphi^2 - s \varphi
\left( 3\, m^2 h + T + \frac{2}{m^2} \p_{\mu}\p_{\nu}T^{\mu\nu}
\right)  + h_{\mu\nu} T^{\mu\nu} \quad. \label{lm1} \eea

\no We can define the generating function

\be Z_{M1}[T] = \int {\cal D}h_{\mu\nu} {\cal D}\varphi {\cal
D}A_{\mu} {\cal D}B_{\mu\nu} \,   e^{i \int d^4x \, {\cal L}_{M1}
} \quad . \label{zm} \ee

\no If we functionally integrate over the extra field $B_{\mu\nu}$
in (\ref{zm}) and reverse the  shift (\ref{sh1}), we come back to
the massive FP theory with the source term we have started with,
namely (\ref{lb}). On the other hand, if we integrate over
$\varphi$ and $A_{\mu}$  in first place we obtain\footnote{We can
always assume $b \ne 6\, s^2$.} the Lagrangian

%\be Z_M[T] = \int {\cal D}h_{\mu\nu} {\cal D}B_{\mu\nu} \,   e^{i \int d^4x \, {\cal L}\left(h_{\mu\nu},B,T}
%\right)} \quad . \label{zm2} \ee
%

\bea {\cal L}(s,b) &=& {\cal L}_{FP}[h_{\mu\nu}] + \frac{m^2}2 B_{\mu\nu}^2  - \frac 13 \left(
\p^{\alpha}B_{\alpha\mu} + \p^{\alpha}h_{\alpha\mu} - \p_{\mu} h - \frac{\p^{\alpha}T_{\alpha\mu}}{m^2}
\right)^2 \nn \\ &-& \frac{s^2}{4\, m^2(6\, s^2 - b)}\left( 3 \, m^2 \, h + T + \frac
2{m^2}\p_{\mu}\p_{\nu}T^{\mu\nu} \right)^2 \quad . \label{lsb} \eea

\no The arbitrariness appears in front of the mass term
proportional to $m^2 h^2$ as in the nFP family (\ref{lnfp}).
Indeed, defining $e_{\mu\nu} = h_{\mu\nu} + B_{\mu\nu} $, the
Lagrangian ${\cal L}(s,b)$ can be rewritten as

\be {\cal L}(s,b) = {\cal L}_{nFP}(c) + h^*_{\mu\nu}T^{\mu\nu} + {\cal O}(T^2) \quad .  \label{lm3}\ee

\no Where ${\cal O}(T^2)$ stands for quadratic terms in the source
and the dual field $h^*_{\mu\nu}$ is given by

\be h^*_{\mu\nu} = e_{(\mu\nu)} - \frac{1+c}3 \eta_{\mu\nu} e - \frac{2(1+c)}{3\, m^2} \p_{\mu}\p_{\nu}e   -
\frac{\left( \p_{\mu}\p^{\alpha}e_{\alpha\nu} + \p_{\nu}\p^{\alpha}e_{\alpha\mu}\right) }{3\, m^2} \label{hdual}
\ee

\no with the arbitrary parameter $c$  defined through

\be b(1+c) = 6 \, s^2(c+1/4) \quad . \label{bc} \ee

\no Comparing (\ref{lm3}) with (\ref{lb}) we conclude that the
master Lagrangian (\ref{lm1}) interpolates between the symmetric
massive FP theory (\ref{lfpd})  and the second family of
one-parameter models nFP given in (\ref{lnfp}). Moreover,
correlation functions of $h_{\mu\nu}$ on the FP side are mapped
into correlation functions of the dual field $h^*_{\mu\nu}$ on the
dual nFP side, up to contact terms, such that we have the dual map
$\left(h_{\mu\nu}\right)_{FP} \leftrightarrow
\left(h_{\mu\nu}^*\right)_{nFP}$.

Since on the nFP side we have both symmetric $h_{\mu\nu}$ and
antisymmetric $B_{\mu\nu}$ tensors, one might ask what is the dual
of  $B_{\mu\nu}$ on the FP side. If we add a source term
$J_{\mu\nu}B^{\mu\nu}$ to the master Lagrangian (\ref{lm1}), the
reader can ckeck that correlations functions of $B_{\mu\nu}$
vanish up to contact terms. This is not surprising, since in the
nFP theory we have on shell $e_{[\mu\nu]}=B_{\mu\nu}=0$ and
equations of motion are enforced at quantum level in the
correlation functions up to contact terms\footnote{This follows
from the functional integral of a total derivative $ \int {\cal
D}B \frac{\delta}{\delta \, B(x)} \left\lbrack e^{i \, \int d^4x B
\, \hat{K} \, B}B(x_1) \cdots B(x_N) \right\rbrack =0 $}.

\no Likewise, we have on shell  $\eta^{\mu\nu}e_{\mu\nu} = h =
0=\p^{\alpha}e_{\alpha\beta}$ which completes the FP conditions
(\ref{c1})-(\ref{c3}). Therefore, up to contact terms, we see from
(\ref{hdual}) that $h_{\mu\nu}$ in the FP theory is mapped simply
into $e_{(\mu\nu)}$ in the nFP family.

Last, we remark that if $b=0$, the arbitrary parameter $s$
disappears from (\ref{lsb}) and we end up with the traceless nFP
theory with $c=-1/4$, see (\ref{bc}). So the arbitrariness of the
nFP family stems indeed from the arbitrary mass term in (\ref{lb})
and not from the arbitrariness in the shift (\ref{sh1}). In the
next subsection we use a fixed shift.

\subsection{Vector Spectator}

Next we interconnect the third family (\ref{la1}) with the usual
massive FP theory. Inspired by (\ref{lfpd}) and (\ref{lb}) we add
an arbitrary mass term for a vector field to the symmetric FP
theory:

\be {\cal L}_{\tilde{b}} = {\cal L}_{FP}[h_{\alpha\beta}] + \tilde{b} \, \frac{m^2}2  \, A^{\mu}A_{\mu} +
h_{\alpha\beta}T^{\alpha\beta} \quad . \label{lbt} \ee

\no After the shift

\be h_{\mu\nu} \to h_{\mu\nu} + (\p_{\mu}A_{\nu} + \p_{\nu}
A_{\mu} )/m \quad . \label{sh3} \ee

\no we obtain a Maxwell term, see (\ref{i2}), which can be brought
to first order again via an antisymmetric field $B_{\mu\nu}$, such
that we derive from (\ref{lbt}) the master Lagrangian

\bea {\cal L}_{M2} &=& {\cal L}_{FP}[h_{\alpha\beta}] + \tilde{b}\, \frac{ m^2}2
\, A^{\mu}A_{\mu} + \frac {m^2}2 B_{\mu\nu}^2 + \, h_{\alpha\beta}T^{\alpha\beta} \nn\\
&+& 2\, m \, A^{\mu}\left( \p^{\alpha}B_{\alpha\mu} +
\p^{\alpha}h_{\alpha\mu} - \p_{\mu} h - \frac 1{m^2}
\p^{\alpha}T_{\alpha\mu} \right) \label{lm2} \eea

\no On one hand, if we integrate over $B_{\mu\nu}$ and reverse the shift (\ref{sh3})
we return to our starting
point (\ref{lbt}). On the other hand, integrating over $A_{\mu}$ in first place we deduce

\be {\cal L}_{M2} = {\cal L}_{FP}[h_{\alpha\beta}]  + \frac{m^2}2
B_{\mu\nu}^2 - \frac 2{\tilde{b}} \left(\p^{\alpha}B_{\alpha\mu} +
\p^{\alpha}h_{\alpha\mu} - \p_{\mu} h \right)^2  \label{lm2b} \ee

\no Defining once again $ e_{\mu\nu} = h_{\mu\nu} + B_{\mu\nu} $
and identifying $\tilde{b}=-2/(a_1-1/4)$ we
rewrite\footnote{Recall that at $a_1=1/4$ the third family
(\ref{la1}) becomes the massive FP theory, so we can assume
without loss of generality $a_1 \ne 1/4$} ${\cal L}_{M2}$ in the
form of the third family (\ref{la1}):

\be {\cal L}_{M2} = {\cal L}_{a_1}[e_{\alpha\beta}]  +
\tilde{h}_{\mu\nu}T^{\mu\nu} + {\cal O}(T^2) \label{lm2c}
\ee

\no where

\be \tilde{h}_{\mu\nu} = e_{(\mu\nu)} + \left(\frac 14 - a_1\right)\left\lbrack
\left(\p_{\mu}\p^{\alpha}e_{\alpha\nu} + \p_{\nu}\p^{\alpha}e_{\alpha\mu}\right) - 2\, \p_{\mu}\p_{\nu} e
\right\rbrack  \quad . \label{ht} \ee

\no We conclude that the master action (\ref{lm2}) interpolates
between the usual massive FP theory, see (\ref{lbt}), and ${\cal
L}_{a_1}$. Since the equations of motion of ${\cal L}_{a_1}$ lead
to $\p^{\alpha}e_{\alpha\mu} = 0 = e = e_{[\mu\nu]}$, all such
terms have vanishing correlation functions up to contact terms. So
we have from (\ref{ht}) the simple map
$\left(h_{\mu\nu}\right)_{FP} \leftrightarrow
\left(e_{(\mu\nu)}\right)_{a_1} $.

Regarding the introduction of interactions, if we had nonlinear
self-interacting terms ${\cal L}_{SI}[h_{\mu\nu}]$ in (\ref{lb}),
after the shift (\ref{sh1}) we would have some nonlinear $\varphi$
dependence in ${\cal L}_{SI}[h_{\mu\nu} - s \,
\eta_{\mu\nu}\varphi -(2\, s/m^2) \p_{\mu}\p_{\nu}\varphi]$. There
is no reason ${\it a}$ ${\it priori}$  for the self-interaction to
be invariant under those spin-0 transformations. Similarly, the
shift $h_{\mu\nu} \to h_{\mu\nu} -s(\p_{\mu}A_{\nu} +
\p_{\nu}A_{\mu})/m $ would lead to some nonlinear $A_{\mu}$
dependence since we do not expect linearized reparametrization
invariance for the full nonlinear theory. Of course, we would
still be able to introduce an antisymmetric field in order to
bring the Maxwell term to first order. However, the nonlinear
terms in $\varphi$ and $A_{\mu}$ in the master action would lead
to a nonlocal dual model after their functional integrals. A
similar conclusion, see (\ref{sh3}), is drawn for the second case
(\ref{lbt}).

\section{Conclusion}

With the help of spectator fields we have been able to
interconnect via the master theories (\ref{lm1}) and (\ref{lm2})
the new one-parameter families of massive spin-2 models
(\ref{lnfp}) and (\ref{la1}) with the symmetric massive
Fierz-Pauli theory (\ref{lfp}). Our master actions offer an
alternative proof of equivalence of the new models, which use a
nonsymmetric tensor, with the fully symmetric FP theory. They
confirm the results of \cite{ms,nfp,rank2}  regarding the
existence of other ghost-free second-order models different from
the FP theory, contrary to early works \cite{rivers}-\cite{cko}.

We have remarked that nonlinear massive gravity models based on
the usual FP theory are not expected to be equivalent to possible
local nonlinear completions of the new models. The situation is
similar to the duality between the second order abelian
Maxwell-Chern-Simons  theory \cite{djt} and the first order
self-dual model of \cite{tpn}. Both models describe an helicity
$+1$ (or $-1$) mode in $D=2+1$. Although there is a master action
\cite{dj} relating those abelian (quadratic) models, the duality
does not go through their non abelian (nonlinear) counterparts due
to extra nonlocal terms, see a discussion in \cite{botta-na}.

The next step is to consider nonlinear (self-interacting)
completions of the new families (\ref{lnfp}) and (\ref{la1}) with
a correct counting of degrees of freedom as expected for a massive
spin-2 particle. Eventually, consistency of the self-interacting
theory may fix the arbitrary parameters $c$ and $a_1$ in
(\ref{lnfp}) and (\ref{la1}). For the FP family (\ref{lfpd}) a
Stueckelberg-like approach \cite{zino} has led to $d_-=1$. In
\cite{bfmrt}  one finds further evidence in favor of $d_-=1$,
since the linearized new massive gravity in $3D$ \cite{bht} and
$4D$ \cite{bfmrt} can be directly (at action level) deduced from
${\cal L}_{FP}(d_-=1)$ by a derivative field redefinition which
holds even at coinciding points (no contact terms). For instance,
one can choose , see \cite{nmgd},
 $e_{\mu\nu}=\p^{\rho}\Omega_{\mu\nu\rho}$, where the
mixed symmetry tensor $\Omega_{\mu\nu\rho}=-\Omega_{\mu\rho\nu}$
is traceless $\eta^{\mu\nu}\Omega_{\mu\nu\rho}=0$.

Finally, since each of the three dual families is obtained by the
addition of a different (less components) kind of spectator field
$B_{\mu\nu}$, $\varphi$, $A_{\mu}$ appearing in
(\ref{lfpd}),(\ref{lb}) and (\ref{lbt}), it is expected that the
approach used here could be generalized in order to find dual
spin-S models not necessarily described by fully symmetric rank-S
tensors $h_{\mu_1, \cdots , \mu_S}$.

\section{Acknowledgements}

We thank Prof. S. Deser for a comment on our previous work
\cite{rank2} which has partially motivated the present note. This
work is supported by CNPq and FAPESP (2013/00653-4).

\end{document}